\documentclass[twocolumn,preprintnumbers,amsmath,amssymb]{revtex4}
\usepackage{graphicx,epsfig}
\usepackage{dcolumn}
\usepackage{bm}
\begin{document}

\title {Bridging the Gap Between the Mode Coupling and the Random First
Order Transition Theories of Structural Relaxation in Liquids}

\author{ Sarika Maitra Bhattacharyya$^{\dagger}\footnote
{Electronic mail~:sarika@sscu.iisc.ernet.in}$, 
Biman Bagchi$^{\dagger}$\footnote{Electronic mail~:bbagchi@sscu.iisc.ernet.in} \\
and \\
 Peter G. Wolynes$^{\ddagger}$\footnote
{Electronic mail~:pwolynes@chem.ucsd.edu}}

\affiliation{$^{\dagger}$ Solid State and Structural Chemistry Unit, 
Indian Institute of Science, Bangalore 560 012, India.\\
$^{\ddagger}$Department of Chemistry and Biochemistry, 
University of California at San 
Diego, La Jolla, California 92093-0371}

\begin{abstract}
A unified treatment of structural
relaxation in a deeply supercooled glassy liquid 
is developed which extends the existing 
mode coupling theory (MCT) 
by incorporating the effects of activated
events by using the concepts from the
random first order transition (RFOT) theory. We show how 
the decay of the dynamic
structure factor is modified by localized activated events 
(called instantons) which lead to the spatial reorganization of molecules
in the region where the instanton pops up. 
The instanton vertex added to the usual MCT depicts the 
probability and consequences of such an event 
which can be derived from the random first
order transition theory. The vertex is proportional to $exp(-A/s_{c})$ where $s_{c}$ 
is the configurational entropy.
Close to the glass transition temperature, $T_{g}$, 
since $s_{c}$ is diminishing, the activated process
slows beyond the time window and this eventually leads to an arrest of the 
structural relaxation as expected for glasses. The combined treatment describes 
the dynamic structure factor in deeply supercooled liquid fairly well, with 
a hopping dominated decay following the MCT plateau.
\end{abstract} 

\maketitle
Inelastic neutron scattering has given detailed structural information on 
how transport in liquids changes as they are cooled in the glassy state 
\cite{book, fayer}. Many
of the structural details are well described by mode coupling theory (MCT)
but 
that approach does not describe well the dynamics of 
the deeply supercooled state 
\cite{rajesh25,leu,beng,gotze,sjogren,sarirev}. 
The random first order transition (RFOT) theory of glass explains this 
lacuna of MCT \cite{xiawoly,Lubwoly,lubwoly2}. 
In this paper we show how RFOT theory can bridge the gap between the 
onset of glassy dynamics at high temperatures and the low temperature behavior 
and thus provide a unified view of the 
structural dynamics in deeply supercooled region.
Over the last few decades, experimental and theoretical  
studies on the phenomenological aspects of
the glass transition have been the subject of a vast scientific literature.
A glass forming liquid is characterized by its ability to circumvent 
crystallization on rapid cooling to a temperature well below its freezing 
temperature $T_{m}$.
In contrast to crystal formation what is most remarkable about solidification
into an amorphous state is that the structure
 changes accompanying vitrification
are either very small or unobservable. 
However, on passage through supercooled regime the 
dynamic and thermodynamic behavior of the liquid
exhibits a number of anomalies as it eventually transforms to an
amorphous solid, called a glass.

The density-density correlation function at finite wave vector probed by 
neutron scattering, $\phi_{k}(t)$ for a system 
approaching glass transition first 
decays via a fast microscopic process which is followed by a plateau in the 
intermediate time. The dynamics while approaching the plateau is 
non-exponential, power law in nature and is called $\beta$ relaxation. 
The time scale of the $\beta$ relaxation is predicted to show a 
power law divergence. The 
decay from the plateau known as the initial $\alpha$ relaxation 
is also given by a power law, known as von Schweidler law. The 
exponent in the $\beta$ regime and the initial $\alpha$ relaxation 
are different in value but related to each other. The slow $\alpha$ 
relaxation which appears at long times is described by an 
stretched exponential or Kohlrausch-Williams-Watts (KWW) 
function \cite{fayer}.

The thermodynamic and kinetic anomalies at $T_{g}$ are known experimentally 
to be correlated.
One experimentally finds a sharp rise in the measured heat capacity of a liquid
during heating which follows prior cooling at a constant rate\cite{heatcp}, 
the cycle being well extended on either sides of the glass transition region.
The over shoot of the heat capacity is taken to be a thermodynamic 
signature of a glass
to liquid transition. However, what lies at the heart of glass 
transition is the 
dramatic slowing down of molecular motion that occurs progressively 
on cooling after the plateau in the inelastic structure factor appears.
This is manifested by a phenomenal increase in shear viscosity $\eta$ 
(also the characteristic structural relaxation time $\tau$) by several 
orders of magnitude for a relatively modest decrease of temperature
(by few tens of degrees). The glass transition seen in the laboratory 
is in fact best described
as a kinetic phenomenon which marks the falling out of the equilibrium 
due to the inability of molecular rearrangement within the experimental 
timescale as temperature T is lowered. A conventional definition 
of the transition temperature $T_{g}$ is the temperature 
at which the viscosity, $\eta$ 
reaches a value $10^{13} P$.

The mode coupling theory makes detailed predictions of the 
temperature 
dependence of the time evolution of the density-density correlation 
function \cite{rajesh25}.
The emergence of a plateau in the neutron scattering 
is one of the important predictions of the ideal mode coupling theory. 
At a quantitative level both the emergence of a plateau in the dynamic 
structure factor at high temperature and the exponential slowing of 
the dynamics 
at low temperature are also explained by the random first order transition 
(RFOT) theory \cite{xiawoly,Lubwoly}. 
In the high temperature limit, RFOT contains the essential elements of MCT as a
stability limit.
While idealized MCT can also correctly predict the power law dynamics in 
the $\beta$ and the initial $\alpha$ 
relaxation regime and the relation between the two exponents in these 
two regimes, it fails to explain the slow $\alpha$ relaxation 
which appears at long times. MCT strictly applied, predicts 
a premature arrest of 
the decay of the density-density correlation function. 
The dynamical plateau of MCT was contemporaneously obtained 
by a variety of static
approaches based on self-consistent phonon calculations and density functional 
theory which form the origin of the RFOT theory \cite{stossel1,stossel}. 
MCT would appear to predict a dynamic arrest of the 
liquid structural relaxation, arising solely from a nonlinear 
feedback mechanism on the memory function. This feedback mechanism 
was first proposed 
by Geszti \cite{ges}
to describe the growth of viscosity in a previtrification region 
and involves a self-consistent calculation of the dynamical correlation 
function and transport coefficients. In RFOT theory one sees the dynamic 
arrest is connected with the emergence of a mean field free energy landscape.

According to the MCT, the essential driving force for glass 
transition is the slowing down of the 
density fluctuations near the wavenumber at $q\simeq q_{m}$ where the static 
structure factor is sharply peaked. Thus, prima facie the glass transition does not appear 
to be caused 
by a small wavenumber infra red singularity observable in the static 
structure factor, but rather it is a phenomenon 
where the intermediate wavenumbers are important, (although a diverging 
length associated with the four point correlations is expected in the mean 
field RFOT theory \cite{kw1987,biroli}). There is a softening 
of the heat mode near $q \simeq q_{m}$.
Kirkpatrick and Wolynes \cite{kw1987} showed that this mode coupling 
transition was equivalent to the transition to an aperiodic crystal 
predicted by the quasi-static theories.  Near the dynamical transition there 
is a general slowing down of all the dynamical quantities, 
and this slowing down 
is most effectively coupled to wavenumbers near $q_{m}$. 
The glass transition temperature predicted by MCT always exceeds the 
laboratory $T_{g}$. 
The explanation for this was pointed out by Kirkpatrick and 
Wolynes \cite{kw1987}: Since there are exponentially many aperiodic structures 
any single one of them is unstable to transformation to the other. 
At $T_{c}$ the system is not frozen because activated events allow 
interconnection between these structures. 
 Thus $T_{c}$ can be identified as a crossover temperature below which
activated motions occur in which groups of particles move in a 
cooperative way over local barriers in the free energy landscape.
Computer simulation studies have amply demonstrated this change of transport 
mechanism and have explicitly revealed the presence of such 
hopping motions in deeply supercooled liquids when the usual convective 
diffusive 
motions cease to exist \cite{wang,arnab}. Simulations also  
show that these hopping events relax the local stress in the system 
\cite{anistress}.

These properties of the activated events below $T_{c}$ can be represented 
by the random first order transition theory in a way superficially analogous 
to the theory of nucleation at random first order transition (although there 
are some differences).
Calculation of fluctuations and transport  
in deeply supercooled or
glassy medium are most easily made where the reference is an 
ideal glassy state and where the configurational entropy has vanished 
\cite{xiawoly,kw,ktw}. The barriers for activated dynamics vanish as the 
temperature of the system approaches the dynamical transition temperature 
from below much like at a spinodal. 
The barrier for activated events can be described as
a nucleation phenomenon where the free energy change is a sum of 
the entropy change 
due to the formation of an entropy droplet and a surface term with a radius
dependent surface energy \cite{xiawoly,Lubwoly}. 
A more explicit view of the similarities and the differences from nucleation 
appears in the recent work of Lubchenko and Wolynes and the elegant analysis 
of Biroli and Bouchaud \cite{biroli}.
Deep in the supercooled regime, the nucleation free energy barrier is given 
by the following simple expression
\begin{equation}
\Delta F^{\star}/k_{B}T = 32 k_{B}/s_{c} \label{barrier}
\end{equation}
where $s_{c}$ is the configuration entropy of a single moving bead i.e 
equivalent spherical particle. 
The coefficient $32$ arises from a specific microscopic calculation and 
reflects the entropy cost of localizing a particle to vibrate in the cage, 
explicitly observed in the plateau of the dynamic structure factor.
If one approximates
$s_{c}$ near the 
Kauzmann temperature $T_{K}$ by $s_{c}=\Delta c_{P} (T-T_{K})/T_{K}$,
one then obtains the empirical Vogel-Fulcher-Tammann (VFT) law, with a relation
between the liquid's fragility and the heat capacity jump, $\Delta c_{P}$.
This expression agrees well with experimental results.
However, Eq.\ref{barrier} is an asymptotic form of the barrier, valid only
near 
$T_{K}$ and as will be discussed later, if we take into account the 
barrier softening effect near $T_{c}$ then the value of the barrier 
will be considerably 
smaller than that given by Eq.\ref{barrier} \cite{Lubwoly}. This effect is 
found to be stronger as we move away from $T_{K}$. 
Once the estimation of the barrier height is made the relaxation time 
$\tau$ is calculated using, $\tau/\tau_{0}=exp(F^{*}/k_{B}T)$. 
The random first order phase transition theory 
also provides an explanation of the dominant source of non-exponentiality 
in relaxation 
dynamics in terms of
distribution of the configuration entropy of the activated droplet. 
If one assumes a Gaussian distribution of the barrier, then the standard 
deviation of the barrier height at $T_{g}$ also turns out to be 
related to the specific 
heat discontinuity of the supercooled liquid. This correlation between 
nonexponentiality and $\Delta C_{p}$ is observed 
in the laboratory \cite{angell}.

In this article we use the concepts of mode coupling theory and random 
first order transition theory to include both the convective diffusional motion
(given by MCT) and the hopping motion (described by RFOT) in the dynamics.
Thus the present treatment also correlates the kinetic anomalies 
(like stretching of the relaxation time) with the thermodynamic 
anomalies (jump in the heat capacity) primarily through the RFOT theory. 
Here we first discuss the effect of hopping alone on the structural 
density fluctuations. 
The rate of hopping and the size of the region involved in hopping are 
calculated from
RFOT theory \cite{Lubwoly}. It is shown that hopping opens up a new channel 
for the structural relaxation.
Next we present the equation of motion of the density-density correlation 
function where the diffusional and the hopping motions are considered as 
two parallel channels for the structural relaxation in the liquid. In the same 
spirit as in the case of the idealized MCT the new 
equation of motion which now has hopping motion in it, 
is calculated self-consistently with the longitudinal 
viscosity. The feedback mechanism which gave rise to the divergence of the 
viscosity in the idealized MCT \cite{leu} 
is also present in our theory but due to the 
presence of hopping, the divergence is shifted to a lower temperature where 
the RFOT hopping rate now finally vanishes only 
as the configurational entropy goes to zero. 
In our proposed equation of motion the diffusional and the hopping motions are 
considered as parallel channels for the density relaxation.
In the first approximate scheme we write the total density correlation 
in the time plane and in the second approximate scheme 
it is written in the frequency plane. We show that not only both 
the approximations (although they appear to be quite different)
lead to similar 
effects on the density-density correlation function, although.
the predicted longtime decays are somewhat different.

Before discussing our own theoretical scheme we would like to mention that 
Das and Mazenko \cite{das} in their fluctuating nonlinear 
hydrodynamic theory  and later Gotze and Sjogren \cite{gotze,sjogren} 
in their extended MCT 
have elegantly demonstrated that in the relaxation kernel (memory function), in addition to the density contribution,
if a coupling to the current modes is also included, 
then below $T_{c}$, 
the dynamic structure factor  decays from the plateau value in the long time.
Gotze and Sjogren \cite{gotze,sjogren}  have 
shown that this modification eventually leads to a parallel decay 
channel of the dynamic structure factor which eliminates the ergodic to 
non-ergodic transition at $T_{c}$. However, although the authors call this 
extra 
decay channel a hopping mode, there is no microscopic connection made in 
those theories to the 
single particle or collective hopping or 
the system crossing over a free energy barrier in the 
free energy landscape. In those treatments, 
the strength and the timescale of this so 
called 
hopping mode was determined by fitting to experimental results, and an
Arrhenius temperature dependence of the rate of "hopping" was assumed
in this fitting.

\section{Theoretical Scheme}
We propose a unified structural description 
covering the whole temperature regime, from the 
high temperature microscopic dynamics described by collisional physics 
to the low temperature collective many particle activated dynamics.
The mode coupling theory analyzes liquid state dynamics by starting 
from a gas-like mode of transport but including correlated collisions and 
back flow. MCT does a good job describing 
structural correlations observed by neutron scattering in the normal 
liquid regime. MCT can also describe the dynamics approaching 
the supercooled regime but fails below $T_{C}$, 
where activated processes also contribute substantially. 
A deeply supercooled liquid explores a very rugged underlying free energy 
surface. The dynamics 
within each well may still be described by the idealized MCT 
which does not take 
into account the possibility of hopping between wells.
Activated events represent essential singularities in the noise strength 
and thus are not contained in the perturbative MCT. 
To include the effect of activated events
on dynamical structure factor, we need to calculate two different
quantities. First, we need to calculate the effect  
such an event (which we call "popping up of an instanton") would have on the
dynamic structure factor. The impact of such an instanton depends
both on its size, that is, how many molecules are involved in the 
activated event and 
on how far particles are displaced by an activated transition. 
In RFOT theory, these quantities are the dynamical correlation length 
$\xi$ and the Lindemann length $d_{L}$.
Second, we also need to know the
probability of such an activated event occurring. 
We can estimate both this probability and the other quantities from 
the random first order transition theory using thermodynamic and structural 
information alone.

A similar underlying physical picture to that presented here has been 
put forward by Fuchizaki and Kawasaki \cite{kawasaki} using what they 
called dynamic density functional theory 
(DDFT) to go beyond the MCT.
In this coarse-grained description one considers the time evolution
of the density field on a lattice. 
The dynamics is supposed to be given 
by a master equation and the free energy functional is taken from 
Ramakrishna-Yussouff (RY) theory \cite{ry}. As shown by Singh, Stoessel and 
Wolynes \cite{stossel}, RY theory can produce multiple minima.
Fuchizaki and Kawasaki have shown that 
transitions between such minima allow
the 
structural relaxation of the coarse grained 
density field. They show these transitions give rise to a stretched 
exponential relaxation. 
In their calculation, 
the overlap between the time 
dependent density field and the reference one of the same lattice site, 
takes a constant 
value, close to unity, in the liquid
but in the supercooled state, the overlap is less than 
one and decreases via discrete jumps which they have connected to activated 
hopping transition over free-energy barriers. They have
 further defined an order parameter which is the overlap of the 
density fields at different lattice sites. The distribution of the order 
parameter, in the normal liquid regime is sharply peaked around unity 
but at higher density the distribution is broad and peaked around a value
less than unity. The broad distribution reflects the multi-minima 
aspects of the free-energy landscape. 

The scheme presented by FK was implemented numerically and 
does not give as explicit picture of the microscopic 
mechanism of hopping as the current treatment does.
 Moreover, due to the temporal coarse graining the 
initial decay of the dynamic structure factor, $\phi(t)$ 
and the subsequent plateau are missed in their analysis. In addition, 
their analysis did not highlight the 
connection with the
configurational entropy of the system.
In contrast the present unified theoretical treatment provides a $\phi(t)$ 
which produces a correct behavior over the full 
time and temperature plane, and also provides an analytical expression of 
the dynamic structure factor where the hopping alone is considered from a 
microscopic point of view. In the following section, we 
find $\phi_{hop}$ or the effect of hopping on the density relaxation. 
Subsequently we incorporate $\phi_{hop}(t)$ in the total $\phi(t)$ 
in a self-consistent way.

\section{Effect of hopping on density relaxation}
Here we describe the hopping in terms of the popping up of a single 
instanton which is an 
activated event whose probability is obtained from the random first order 
transition theory.
Let us say that an instanton pops up at a position $R$. 
We will take this region of influence to be spherical on the average. 
Near $T_{A}$, good arguments suggest the region actually 
is fractal or string-like \cite{wolyunpub}. But the effect will ultimately 
be spherically averaged anyway 
as we shall see below. 
The particles within the sphere of radius $\xi$ around the position $R$
will be displaced by a Lindemann length, $d_{l}$.
Now the change in density $\rho(r)$ due to a 
single instanton popping up can be written as,
\begin{eqnarray}
\delta \rho^{new}({\bf r},t+\delta t)&=& \delta \rho({\bf r},t )+
\theta ((r-R)<\xi)\nonumber\\
&&
\times \Biggl[
\int dt^{\prime}
\int_{{\cal D}(R)} 
d{\bf r}^{\prime} \:\delta \rho({\bf r,}^{\prime}, t^{\prime}) \nonumber\\
&&~~~~~ 
\times G ({\bf r},{\bf r}^{\prime},t-t^{\prime}) 
\theta((r^{\prime}-R)<\xi)\nonumber\\ 
&&~~~~~~~~~~~~~~~~~~~~~~~~~~~~~
- \delta \rho ({\bf r},t) \Biggr]
\end{eqnarray}
\noindent
Where $\theta(({\bf r} -{\bf R})<\xi)$ provides the effect of the 
instanton felt at the position ${\bf r}$, provided it 
is within the radius $\xi$. $G ({\bf r}, {\bf r}^{\prime}t-t^{\prime})$ is the Greens 
function which determines the effect of instanton 
in moving particles from 
position ${\bf r}$ to a new one ${\bf r}^{\prime}$,
typically Lindemann length 
away during time $t$ and 
$t+\delta t$. ${\cal D}(R)$ determines 
the region where the effect of the instanton is felt.

Since a given instanton pops up at a specific location the transformation 
is not 
translationally invariant, but when averaged over the droplet location 
and orientation we restore translational and rotational invariance, 

\begin{eqnarray}
\overline {\delta {\rho}}^{new}({\bf r},t+\delta t)&=& \delta \rho({\bf r},t) + 
\frac{1}{V}
\int d{\bf R}\: 
\theta (({\bf r}-{\bf R})<\xi) \nonumber\\
&&\times \Biggl[ 
\int dt^{\prime} \int_{{\cal D}(R)} 
d{\bf r}^{\prime} \:\delta \rho({\bf r}^{\prime}, t^{\prime}) \nonumber\\
&&~~~~~~~~~~
\times
G ({\bf r}, {\bf r}^{\prime}, t-t^{\prime}) 
\theta(({\bf r}^{\prime}-{\bf R})<\xi)\nonumber\\
&&~~~~~~~~~~~~~~~~~~~~~~ 
- \delta \rho ({\bf r},t) \Biggr] \nonumber \\
&=& 
\delta \rho({\bf r}, t) +
\frac{1}{V} 
\int dt^{\prime}\int_{{\cal D}(R)} 
d{\bf r}^{\prime} \:\delta \rho({\bf r}^{\prime}, t^{\prime}) \nonumber\\
&&~~~~~~~~~~~~~~~~~
\times 
G ({\bf r}, {\bf r}^{\prime}, t-t^{\prime}) 
\Omega({\bf r}^{\prime}-{\bf r}) \nonumber\\
&&~~~~~~~~~~~~~~~~~~~~~~~~~~~~~
- \frac {v_{0}}{V} \delta \rho ({\bf r}, t) 
\end{eqnarray}
\noindent
Where,
\begin{eqnarray}
\Omega({\bf r}-{\bf r}^{\prime})&=& \int d{\bf R} 
\:\theta(({\bf r}-{\bf R})<\xi)
\theta(({\bf r}^{\prime}-{\bf R})<\xi) \nonumber\\
&=&\frac{1}{12} \pi (4\xi+|{\bf r}-{\bf r^{\prime}}|)
(2\xi-|{\bf r}-{\bf r^{\prime}}|)^{2}, \nonumber\\
&&~~~~~~~~~~~~~~~~~~~~~~~~~~for 
({\bf r}- {\bf r}^{\prime}) \leq 2\xi \nonumber\\
&=& 0~~~~~~~~~~~~~~~~~~~~~~~~for 
({\bf r}- {\bf r}^{\prime})> 2\xi
\end{eqnarray}
\noindent
is the overlap volume of two spheres of radius $\xi$ centered at $r$ and 
$r^{\prime}$, respectively,and 
\begin{equation}
v_{0}=\int d{\bf R} \: \theta(({\bf r}-{\bf R})<\xi)
=\frac{4}{3} \pi \xi^{3}
\end{equation}
\noindent
is the volume of the region participating in hopping.
$\xi$ defines the region of hopping which is calculated 
from RFOT theory \cite{Lubwoly}.

The RFOT $\xi$ is of the order of a molecular length,'a' at $T_{c}$ but 
diverges as $(T-T_{K})^{-2/3}$ as the Kauzmann temperature is approached. Near 
$T_{g}$ (where the timescale is of the order of 1 hour) $\xi$ is universally
$\sim$ 5 molecule spacings \cite{Lubwoly}.

Thus expanding 
$\overline {\delta {\rho}}^{new}({\bf r},t+\delta t)$ about $t$ we get,

\begin{eqnarray}
\frac {d \overline {\delta {\rho}}^{new}({\bf r},t)}{dt}\times \delta t
&=&\frac{1}{V}\Biggl[ \int dt^{\prime}\int_{{\cal D}(R)} 
d{\bf r}^{\prime} \:\delta \rho({\bf r}^{\prime}, t^{\prime})\nonumber\\
&&~~~~~~\times  
G ({\bf r}, {\bf r}^{\prime}, t-t^{\prime}) 
\Omega({\bf r}^{\prime}-{\bf r})\nonumber\\
&&~~~~~~~~~~~~~~~~~~~~~ 
- v_{0} \delta \rho ({\bf r}, t) \Biggr ]
\end{eqnarray}
\noindent
Now lets say that the rate of a particle hopping is given by $p_{h}$.
Thus the probability of any particle hopping in the region of volume $V$ 
during the time interval 
$\delta t$ is $p_{h}V \times {\delta t}/v_{P}$, where $v_{p}$ is the volume of a 
particle ( a particle will be a  bead according to RFOT theory).
If we include this information in the above equation and take a 
Laplace transform we get,

\begin{eqnarray}
s\overline {\delta \rho ^{new} ({\bf q},s)}&-& 
\overline {\delta \rho ^{new} ({\bf q},t=0)}\nonumber\\
&=& \frac {p_{h}}{v_{p}}\Biggl[ \int d{\bf q}_{1} G({\bf q}_{1}, s) 
\Omega ({\bf q}-{\bf q}_{1}) \delta \rho({\bf q},s)\nonumber\\
&&~~~~~~~~~~~~~~~~~~~ - v_{0} 
\delta \rho({\bf q},s)\Biggr]
\end{eqnarray}

Thus we obtain,
\begin{equation}
\phi_{hop}(q,s)=\frac{\phi_{hop}(q,0)}{s+(v_{0}-{\widehat {\Omega G}}({q},s) 
)p_{h}/v_{p}} \label{phihop8}
\end{equation}
\noindent
In calculation of the hopping rate we have considered that hopping is 
instantaneous. But simulations have shown \cite{anistress} that once it occurs, 
hopping happens over a period of time. This would give a further frequency 
dependence to $\Omega$.
The fluctuations of entropic driving force leads to a distribution of 
hopping rates $p_{h}$ \cite{lubwoly2}. 
To account for this we can consider that there are n-type of 
instanton with lifetimes ($\tau_{j}$) and with occurrence probability 
$p_{h}^{i}$. In this case equation \ref{phihop8} 
can be averaged over different instanton lifetimes and rates. 
Here, for simplification we replace $p_{h}$ by the average hopping 
rate $P$ which is given by $P=\frac{1}{n}\Sigma_{i} p_{h}^{i}$. Since 
hopping rate is small (that is, hops are rare), we can assume that 
the duration or lifetime of a hop (instanton popping up) is 
infinitesimally small.

In the above equation,
\begin{eqnarray} 
{\widehat {\Omega G}}({q}, s)=\int d{\bf q}_{1} G ({\bf q}_{1}, s) 
\Omega ({\bf q}- {\bf q}_{1}) 
\end{eqnarray}
In the above theory the length scale of the Greens function is given by the 
Lindemann length $d_{l}$ where as the length scale of $\Omega$ is given by 
$\xi$. The random first order transition theory gives us the estimate of 
$\xi$, that is the spatial extent of the instanton and also the rate 
at which the instantons pop up that is $p_{h}$ or more properly its distribution. 
 The latter should be related 
to the waiting time distribution of hopping, as discussed by De Michele and
Leporini \cite{lepo} .

Since $\xi >> d_{l}$, $G(q,s)$ has a much slower decay with $q$ than 
$\Omega(q-q_{1})$ which implies that the product decays slowly with $q$.
The propagator $G(q,s)$ should be just the short time part of 
the dynamic structure factor.

Now we can write,
\begin{eqnarray}
\phi_{hop}(q,s)&=&
\frac{1}{s+(v_{0}-{\widehat {\Omega G}}({ q},s) 
)P/v_{p}} \nonumber\\
&=&\frac{1}{s+K_{hop}(s)}
\label{fhop}.
\end{eqnarray}
\noindent

As mentioned earlier, the length
scale of the Greens function is determined by the 
Lindemann length $d_{l}$. If we neglect its frequency dependence then 
the Greens function can be written as,
$G(q) = exp(-q^{2}d_{l}^{2})$. 
\subsection{Calculation of the hopping rate from RFOT theory}
In this section we briefly discuss the calculation of the 
hopping rate $P$  from the RFOT theory.
As has been discussed before, the hopping rate of an instanton
is connected to the free 
energy barrier, which in turn is connected to the
configurational entropy of the system \cite{xiawoly,Lubwoly} 
and an asymptotic form of that is given by Eq.\ref{barrier}. 
Note that, in glasses, we need the interface energy between 
two alternative amorphous packings and not between two thermodynamic
phases, like a solid and a delocalized uniform liquid.  
The alternative phases become more alike at $T_{c}$.
The interfacial energy (surface tension) that enters
in the calculation of the instanton barrier is smaller at $T_{c}$ than at 
lower temperature. 
This effect has been  discussed by Lubchenko and Wolynes who termed it
as barrier softening. A rigorous estimate of the barrier requires complex 
replica instanton calculations as recently outlined by Franz \cite{franz} 
and by Dzero, 
Schmalian and Wolynes \cite{wolycond}. An alternative method motivated by variational reasoning proposed by 
Lubchenko and Wolynes is to write a simple interpolation formula for the free 
energy which allows the system to choose the smallest available barrier,and is
given by \cite{Lubwoly},

\begin{equation}
F(r)= \frac{\Gamma_{K} \Gamma_{A}}{\Gamma_{K}+ \Gamma_{A}} -\frac {4 \pi}{3} 
r^{3} T s_{c}\:, \label{bsoft}
\end{equation} 
\noindent
where $\Gamma_{K}$ and $\Gamma_{A}$ are the surface energy terms at 
$T_{K}$ and $T_{A}$ respectively.  
$T_{A}$ gives an estimate of temperature at which a 
soft crossover to the onset of activated transport takes place or the 
temperature at which the barrier disappears when approached from below. 
The expression for the surface energy at $T_{K}$ is,
$\Gamma_{K}=4 \pi \sigma_{o}a^{2} (r/a)^{3/2}$, where 
$\sigma_{o}=\frac{3}{4}(k_{B}T/a^{2}) ln((a/d_{L})^{2}/\pi e)$ is the 
surface tension coefficient  and $a$ is the molecular bead-size which refers 
to the size of the quasi-spherical parts of any a spherical molecule. The bead 
count 'b' for each molecule is also be calculated within RFOT theory. 
The expression for the surface energy at $T_{A}$ is, $\Gamma_{A}=
4 \pi \Delta f a^{3} (r/a)^{2}$, where 
$\frac{\Delta f}{T_{A}s^{A}_{c}}= 4 \bigl(\frac{t}{t_{K}}\bigr)^{3/2} 
(1/\sqrt{1+3t/t_{K}})$, 
where $t=(1-T/T_{A})$ and $\Delta f$, the excess free energy density computed 
from the density functional. 
Note that Eq.\ref{bsoft} tends to $\Gamma_{K}$ as $
T \rightarrow T_{K}$ and similarly,  to $\Gamma_{A}$ as $T \rightarrow T_{A}$.
From the above expression of the barrier, the critical 
barrier free energy, $F^{*}(r)$, can be calculated as 
the maximum value of $F(r)$.
Microscopic calculation giving $T_{A}$ can then be used with 
Eq.\ref{bsoft} to calculate the value of viscosity,
$\eta/\eta_{0}=exp(F^{\star}/k_{B}T)$, (where $\eta_{0}$ is the viscosity 
at the 
melting temperature, $T_{m}$ and is obtained from experiments). Rather than 
carry out completely microscopic calculations, Lubchenko and Wolynes 
used this approach to fit 
the experimental values of viscosity, where the values of the transition temperature 
$T_{A}$ and $s_{fit}$ 
($s_{c}=s_{fit}(1-T_{K}/T)$) are used as the fitting parameters.
$T_{K}$ is obtained from thermodynamics. For example, 
$T_K$ is known to be 175K for Salol.

For our calculation of the hopping rate $P$, we use Eq.\ref{bsoft} with the 
fitted values of $T_{A}$ and $s_{fit}$ from Lubchenko and Wolynes
\cite{Lubwoly}. 
Thus $P=\frac {1}{\tau_{0}}exp(-\Delta F^{\star}/k_{B}T)$, 
where $\tau_{0}$ is the timescale of the system at the melting point, $T_{m}$ 
which is taken from experimentally known value \cite{stickel}. 
The above expression would predict that rate of hopping 
becomes of the order of hours at $T=T_{g}$ and vanishes at $T=T_{K}$ 
\cite{Lubwoly}.

We also need an estimate of the size of the hopping region $\xi$ which has a temperature 
dependence and is equated to the $r$ value where in Eq.\ref{bsoft}, F(r)=0.

\section{The total density-density time correlation function}

 In the last section we have evaluated the change in the 
dynamic structure factor due to hopping alone.
In this section we incorporate this effect along with mode coupling feedback terms 
in the full 
density-density correlation function in a self-consistent manner.

As mentioned earlier MCT describes well the dynamics for a moderately 
supercooled system. It also works well for a deeply supercooled system confined 
within a single free energy minimum. These features are preserved when the 
hopping term or instanton vertex is added.
In the idealized MCT, the equation of motion for the 
dynamic structure factor,${\bf\phi}^{id}_{MCT}(q,t)$ is first simplified by 
neglecting the coupling between
 different wave vectors and considering the contribution from the static and 
dynamic quantities at a single 
wave number, $q=q_{m}$, ($q_{m}$ is the wavenumber where the peak of the 
structure factor appears). The idealized MCT equation 
can be written as \cite{leu,beng},
\begin{eqnarray}
&&\ddot{\bf\phi}^{id}_{MCT}(t) 
+\gamma \dot{\bf \phi}^{id}_{MCT}(t) 
+ \Omega_{0}^{2} 
{\bf\phi}^{id}_{MCT}(t)\nonumber\\
&&~~+ 4 \lambda \Omega_{0}^{2} 
 \int_{0}^{t} dt^{\prime} {{\bf\phi}^{id}_{MCT}}^{2}(t^{\prime}) 
\dot{\bf\phi}^{id}_{MCT}(t-t^{\prime}) = 0 \label{fqtleu}
\end{eqnarray}
\noindent 
where ${\bf\phi}^{id}_{MCT}$ is the normalized density time correlation function.  
This integro-differential equation 
 has an unusual structure but it can be solved numerically and many of its properties are now 
analytically understood. The frequency of the free oscillator can be 
approximated as, $\Omega_{0}^{2}=k_{B}Tq_{m}^{2}/m S(q_{m})$ 
and the damping constant, $\gamma$, 
which is the short time part of the memory kernel, 
is taken to be proportional to $\Omega_{0}$.
This is because randomizing collisions occur in Lennard-Jones liquids on nearly the same timescale as the 
vibrations. 
The fourth term on the left hand side has the form of a memory kernel 
and its strength is controlled by the dimensionless coupling constant which
can be taken to be,
$\lambda=(q_{m} A^{2}/8 \pi^{2} \rho) S(q_{m})$, where $A \delta(q-q_{m})
=S(q_{m})-1 $ \cite{beng}.
It is useful to transform to 
the Laplace frequency plane, where the Laplace transform is defined as 
$\phi(s)={\cal L} [\phi(t)]$. Now the equation can be rewritten as,

\begin{eqnarray}
{\bf \phi}^{id}_{MCT} (s)=\frac{1}{s+ \frac {\Omega_{0}^{2}}
{s+\eta^{id}_{l}(s)}}\label{fqtmcts}
\end{eqnarray}
\noindent
where the longitudinal viscosity $\eta^{id}_{l}$ is given by 
\begin{eqnarray}
\eta^{id}_{l}(s)=\gamma + 4 \lambda \Omega_{0}^{2} {\cal L} [
{{\bf \phi}^{id}_{MCT}}^{2}(t)]
\label {etal}
\end{eqnarray}
\noindent
 The above equations are nonlinear in nature and when Eq.\ref{fqtleu} 
is solved
 numerically or Eqs.\ref{fqtmcts} and \ref{etal} are solved self consistently 
 for different values of 
$\lambda$, a dynamical arrest of the liquid structural 
relaxation was predicted at $T=T_{c}$ \cite{leu,beng,gotze}. 
However, $T_{c}$ is always higher than $T_{g}$. The origin 
for this inconsistency is the exclusion of the hopping motion 
in the perturbative MCT dynamics.  

Hopping and continuous motion are essentially 
two distinct channels for the structural 
relaxation in a supercooled liquid, although they get coupled due to
self-consistency, as described below. In the deeply supercooled 
regime these two styles of motion 
are 
usually well separated in their timescales. Thus to include hopping in the 
dynamics of structural relaxation, as a first approximation, valid in the 
deeply supercooled regime, the 
full intermediate scattering function can be written as 
a product of two functions \cite{bbhop},
\begin{equation}
\phi(q,t)\simeq \phi_{MCT}(q,t)\phi_{hop}(q,t) \label{fqthop}
\end{equation}
\noindent
where $\phi_{MCT}(q,t)$ is the mode coupling part and $\phi_{hop}(q,t)$ is the 
contribution from the hopping motion, which is already discussed in the last 
subsection.

 We now derive the equation for $\phi_{MCT}(q,t)$ consistent 
with Eq.\ref{fqthop}. To begin with, note that
the equation of motion ${\bf\phi}_{MCT}(t)$ should remain the same as  
${\bf\phi}^{id}_{MCT}(t)$ (Eq.\ref{fqtleu})
but due to the presence of hopping, the memory function will
now be modified. The memory function, which is the longitudinal
viscosity, is now determined by the full dynamic structure factor which
is given by Eq.\ref{fqthop}. Therefore, ${\bf\phi}_{MCT}(t)$ is now solved 
self consistently with the full 
dynamic structure factor, ${\bf\phi}(t)$, and the equation of motion is 
written as,
\begin{eqnarray}
&&\ddot{\bf\phi}_{MCT}(t) 
+\gamma \dot{\bf \phi}_{MCT}(t) 
+ \Omega_{0}^{2}
{\bf\phi}_{MCT}(t)\nonumber\\
&&~~~~+ 4 \lambda \Omega_{0}^{2} 
 \int_{0}^{t} \:dt^{\prime} {\bf\phi}^{2}(t^{\prime}) 
\dot{\bf\phi}_{MCT}(t-t^{\prime}) = 0 \label{fqtmod}
\end{eqnarray}
\noindent 
Note that  for simplicity we have removed explicit $q$ dependence 
in the above equation and all the quantities are calculated at $q=q_{m}$.
Although, the structure of Eq.\ref{fqtleu} and Eq.\ref{fqtmod} are quite 
similar, the memory function in the former is calculated using 
only $\phi^{id}_{MCT}(t)$ whereas in the later it is calculated using the 
full intermediate 
scattering function, $\phi(t)$. Thus ${\phi}_{MCT}(t)$ is now dependent 
on the hopping motion. 
The equation in the Laplace frequency plane, for the modified 
${\bf\phi}_{MCT}(s)$ remains the same as Eq.\ref{fqtmcts}, except that the 
ideal longitudinal viscosity $\eta^{id}_{l}$ is replaced by the following 
longitudinal viscosity given by, 

\begin{eqnarray}
\eta_{l}(s)=\gamma + 4 \lambda \Omega_{0}^{2} {\cal L} [\bf \phi^{2}(t)]
\label {etalmod}
\end{eqnarray}
\noindent 
We see at this order, the longitudinal viscosity is modified 
which in turn modifies the decay of ${\bf\phi}_{MCT}(t)$.

Combining Eq.\ref{fqthop} and Eq.\ref{fqtmod} we can now write a equation 
of motion for the full structure factor, $\phi(t)$. The equation becomes simple when 
$\phi_{hop}(t)$ can be approximated as an exponential with time scale $1/K_{hop}$. 
In the limit of small $K_{hop}$ the equation of motion for $\phi(t)$ can be 
written as,

\begin{eqnarray}
\ddot{\bf\phi}(t) 
&+&\gamma \dot{\bf \phi}(t) 
+ (K_{hop}\gamma+\Omega_{0}^{2})
{\bf\phi}(t)  \nonumber\\ 
&+& 4 \lambda \Omega_{0}^{2} 
 \int_{0}^{t} \:dt^{\prime} \phi_{hop}(t^{\prime}){\bf\phi}^{2}(t^{\prime})
\nonumber\\
&&~~~~~~~\times 
\biggl[\dot{\bf\phi}(t-t^{\prime})+K_{hop}{\bf\phi}(t-t^{\prime})\biggr] = 0 
\label{schm1}
\end{eqnarray}
\noindent 

The effects of hopping motion in the full structural relaxation can be 
incorporated directly in the frequency plane 
which is 
suggested by Eq.\ref{fhop} 
and Eq.\ref{fqtmcts}. Exploiting the strict parallelism 
of hopping and convective motion,
these two equations can be combined 
to write an extended equation for the dynamic 
structure factor in the frequency plane as,

\begin{eqnarray}
\bf \phi (s)&=&\frac{1}{s+
(v_{0}-{\widehat {\Omega G}}({ q_{m}},s) 
)P/v_{p}+
 \frac {\Omega_{0}^{2}}{s+\eta_{l}(s)}}\nonumber\\
&=&\frac{1}{s+K_{hop}(s) + K_{MCT}(s)} \label{fqt2s}
\end{eqnarray}
\noindent
where
\begin{eqnarray}
&K_{hop}(s)=(v_{0}-{\widehat {\Omega G}}({q_{m}},s) 
)P/v_{p} \nonumber\\
&and \nonumber\\
&K_{MCT}(s)= \frac {\Omega_{0}^{2}}{s+\eta_{l}(s)} \label{mem2s}
\end{eqnarray}
\noindent
and $\eta_{l}(s)$ is given by Eq.\ref{etalmod}.

$K_{MCT}(s)$ and $K_{hop}(s)$ are the contributions to the rate 
of relaxation from MCT and hopping
modes, respectively, and they act as two parallel channels for the decay 
of $\phi(t)$.
At high temperature and low density $K_{MCT} >> K_{hop}$, thus the decay of 
$\phi(t)$ is determined primarily by the MCT term. But as we approach 
$T_{c}$, the relaxation of $\phi (t)$ slows down which increases the 
longitudinal viscosity value (given by Eq.\ref{etal}). This in turn 
causes $K_{MCT}(s)$ to decrease much faster than $K_{hop}(s)$ at low
frequency.  
Now if we neglect the frequency dependence of $K_{hop}$ then the equation of 
motion for $\phi(t)$ for small $K_{hop}$ value can be written as,
\begin{eqnarray}
\ddot{\bf\phi}(t) 
&+&\gamma \dot{\bf \phi}(t) 
+ (K_{hop}\gamma+\Omega_{0}^{2})
{\bf\phi}(t)  \nonumber\\
&+& 4 \lambda \Omega_{0}^{2} 
 \int_{0}^{t} \:dt^{\prime}{\bf\phi}^{2}(t^{\prime})\nonumber\\
&&~~~~~\times
\biggl[\dot{\bf\phi}(t-t^{\prime})+K_{hop}{\bf\phi}(t-t^{\prime})\biggr] = 0 
\label{schm2}
\end{eqnarray}
\noindent 

Note that as expected, 
while Eq.\ref{schm2} has similar structure as Eq.\ref{schm1}, 
the memory function is different and the consequence of this will be discussed 
later. 
 
The second scheme (given by Eq.\ref{fqt2s}) appears to be rather similar to 
extended mode coupling theory of Gotze and Sjogren\cite{gotze,sjogren}. 
In the extended MCT the current contribution is incorporated in the memory 
kernel. The modified memory kernel leads to an extra decay channel of
the dynamic structure factor thus successfully 
predicting the decay of $\phi(t)$ 
below $T_{c}$. The rate of decay for the second channel, $\delta$, 
is called the rate due to the hopping mode. 
However, note that there is no direct connection between 
the expression of $\delta$ in the Gotze-Sjogren \cite{gotze} analysis 
and the hopping motion. Since the microscopic 
calculation of $\delta$ can never predict a zero hopping rate they used 
$\delta$ as a fitting parameter to explain the 
experimental results.

\section{Numerical calculations and results}


For this paper we solved
Eqs.\ref{fqtleu}, \ref{fqtmod},\ref{schm1} and \ref{schm2} 
are solved numerically. 
Although in principle Eqs.\ref{fqt2s}-\ref{mem2s} can be solved self 
consistently with Eq.\ref{etalmod} in an 
iterative way, it requires a very large number of iterations
to converge near the glass transition temperature.
An alternative way is to numerically solve directly Eq.\ref{schm2}. 
Both these calculations were done for a model system and for concreteness were applied to a specific 
molecular system, Salol. 

\begin{figure}
\epsfig{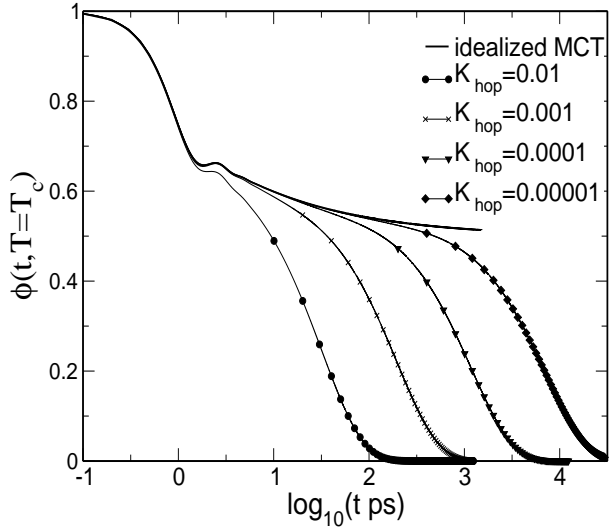}
\caption{The idealized MCT result (given by Eq.\ref{fqtleu}) 
and the modified 
full $\phi(t)$ (given by Eq.\ref{schm1}) 
have been plotted against log(t ps) for different values of $K_{hop}$. 
All the plots are for $\lambda=1$ that is at $T=T_{c}$, the 
mode coupling transition temperature. For different values of 
$K_{hop}$ the curve follows the idealized MCT result till it starts 
decaying from the plateau. The smaller the $K_{hop}$ value the longer 
is the plateau and the slower the long time decay.}
\end{figure}

\begin{figure}
\epsfig{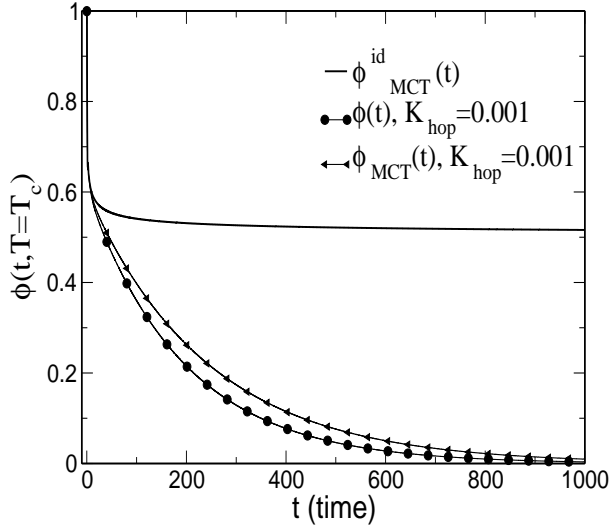}
\caption{The total $\phi(t)$(given by Eq.\ref{schm1}), the modified 
$\phi_{MCT}(t)$(given by Eq.\ref{fqtmod}) and the idealized 
$\phi^{id}_{MCT}(t)$(given by Eq.\ref{fqtleu}) are plotted
against time for $\lambda=1$ ( i.e $T=T_{c}$) and for $K_{hop}=.001$. 
After the initial short time decay, 
both the total structural relaxation $\phi(t)$ and the MCT part of the 
structural relaxation $\phi_{MCT}(t)$ decays with time where as the 
idealized MCT theory predicts no decay of the structural relaxation. 
This figure shows that the  
decay of $\phi_{MCT}(t)$ due to hopping induces 
continuous diffusion in the system. }
\end{figure}

\begin{figure}
\epsfig{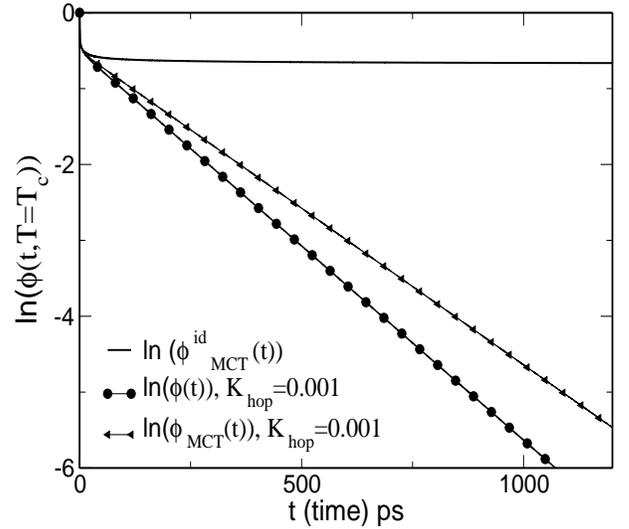}
\caption{$ln(\phi(t))$, $ln(\phi_{MCT}(t))$ and 
$ln(\phi^{id}_{MCT}(t))$ have been plotted against time where the parameters 
are the same as in figure 2. The plot shows that in the longtime 
both $\phi(t)$ and $\phi_{MCT}(t)$ decays exponentially with different 
timescales. Due to the presence of the explicit hopping term 
(see Eq.\ref{fqthop}) $\phi(t)$ decays faster than $\phi_{MCT}(t)$.}
\end{figure}
\subsection{Results for the schematic model}
In the numerical calculations of the schematic model system, 
we took $\Omega_{0}=1$, 
and $\gamma=\Omega_{0}$ fixed. We varied $\lambda$ between .1-1, knowing that $T_{c}$ is 
reached for $\lambda=1$ and varied $K_{hop}$ from 
0.01 to 0.0001. 

In {\bf figure 1} we plot both the idealized MCT result (given by Eq.\ref{fqtleu}) 
and the modified 
full $\phi(t)$ (given by Eq.\ref{fqthop}) 
dynamic structure factor against log(t). 
We see that although the ideal MCT saturates to a plateau value, 
the extended theory exhibits a 
hopping induced decay following the MCT plateau value -- the duration
of the plateau depends on the rate of hopping.

In {\bf figure 2} we plot the total $\phi(t)$ and $\phi_{MCT}(t)$ 
against time.
The plot shows that hopping not only leads to the decay 
of the total $\phi(t)$ (given by Eq.\ref{fqthop}), but 
it also slows down the growth of the 
longitudinal viscosity thus facilitating the structural relaxation 
even through the mode coupling channel.

The long time behavior of $ln(\phi(t))$ and $ln(\phi_{MCT}(t))$ are plotted 
against time in 
{\bf figure 3}, for $\lambda=1$ and $K_{hop}=.001$. 
The semilog plot shows straight lines in the long time with different  
slopes for the MCT part and the total dynamic structure factor, thus  
indicating that both the functions are exponential 
in the long time but with somewhat different time constants 
as evident in figure 2.

Next we compare the two different schemes at $T=T_{c}$ (that is, for 
$\lambda=1$) and for $K_{hop}=.001$. The solution of Eq.\ref{schm1} 
and Eq.\ref{schm2} are plotted in {\bf figure 4}
against log(t). 
Both the approximate schemes provide similar results but the first scheme 
(Eq.\ref{schm1}) shows a faster decay implying that hopping has a stronger 
effect at this level. 
A comparison between Eq.\ref{schm1} and Eq.\ref{schm2} 
shows that although the structures of the equations are similar, in
the memory function of Eq.\ref{schm1} there is an extra product 
with $\phi_{hop}$ which 
eventually leads to a faster decay of the total function as seen in the figure.
The advantage of the first approximate scheme is that 
we can separately  calculate the MCT part of the dynamic structure factor and 
can then demonstrate explicitly how hopping induces a  
decay of $\phi_{MCT}$. This 
implies that hopping opens up continuous diffusion channels in the system.  

In the schematic calculation we have independently varied $\lambda$ and $K_{hop}$ 
while keeping parameters $\Omega_{0}$ and $\gamma$ constant. 
In real systems all of these are 
functions of density and temperature and needs to varied simultaneously.
Although the temperature dependence of $\Omega_{0}$ and $\gamma$ are not too 
strong but $\lambda$ depends more strongly on T and $K_{hop}$ is expected to depend more 
strongly than exponential on $T$. This means that
as we increase the timescale of hopping (which is achieved by lowering the 
temperature), the $\lambda$ value also increases. 
Thus for real systems, unlike as shown in figure 1, 
as $1/K_{hop}$ gets stretched, the MCT plateau value also should 
increase somewhat. This is consistent with density functional 
theory of the aperiodic solid.
In the next subsection we apply our theory to a real system 
chosen as Salol and include these effects.
\begin{figure}
\epsfig{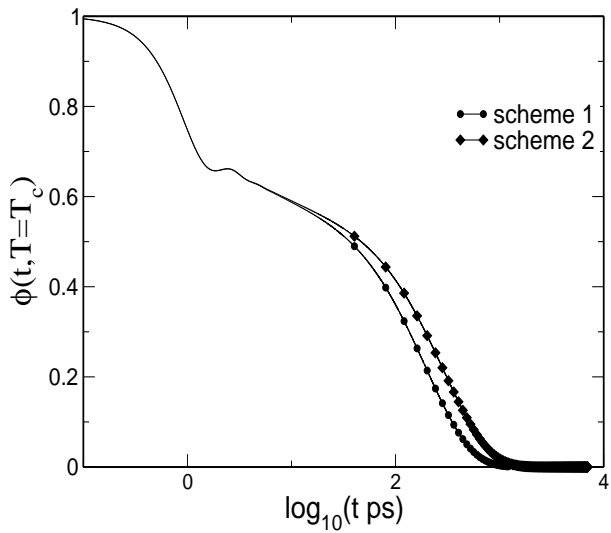}
\caption{To compare the two different 
 approximate schemes, give by Eq.\ref{schm1} (scheme 1) and 
by Eq.\ref{schm2} (scheme 2) we plot $\phi(t)$ 
against log(t ps) for $\lambda=1$ (that is, at $T=T_{c}$) 
and for $K_{hop}=.001$. The initial short time dynamics 
is identical for both the schemes but scheme 1 predicts a slightly 
faster longtime 
decay (see text for detailed discussion).}
\end{figure}

\subsection{Salol}

Salol (phenyl salicylate) is a fragile glass forming substance.
It is intra-molecularly hydrogen bonded, a van der Walls system
which is considered as a model substance for the study of glass
transition and molecular mobility in the supercooled liquid.
To perform the microscopic calculations, we 
 need to map the system 
into a Lennard-Jones system and for that we 
need the molecular diameter and the well depth. 
Although the shape of the molecule is not 
spherical we can calculate a hard sphere diameter $\sigma_{hs}=7.22\AA$ 
\cite{bondi}. Note that in the RFOT theory the elementary 
particles in the system 
are not the molecules but  beads and each molecule is made up of certain 
number of beads. We thus need to estimate the bead size {\it a}. 
Use of the RFOT theory (with softening) for the fit of the 
viscosity leads to 
the number of beads in a Salol  molecule to be 6.29 \cite{Lubwoly}.  
Equating the volume of the molecule with the volume of 6.29 beads,
we get the hard sphere bead size 
$a_{hs}=3.095\AA$ which 
for the present calculation we equate to the Lennard-Jones bead size {\it a}.   
Next we need to estimate the temperature 
scaling which will be equivalent to the Lennard-Jones well depth $\epsilon$.  
To obtain the temperature scaling we use the fact that for
 reduced Lennard-Jones system the mode coupling transition is known to 
take place at 
reduced density, $\rho^{*}=0.95$ and reduced temperature, $T^{*}=0.57$. 
We are not aware of any reported density value of the Salol system, 
but a reduced 
temperature $T^{*}_{c}=0.57$ seems to be a reasonable value when compared with 
simulation results \cite{arnab}. 
For a Salol system it is known that the mode coupling 
transition temperature, $T_{c}=256 K$ 
\cite{stickel25}. Thus the temperature is to be scaled by,
$\epsilon/k_{B}=449.122$. 

\begin{table}
\caption [Table n:] { In this table we present the value of the parameters 
calculated. $\Omega_{0}$ is the frequency of the free oscillator, $\gamma$ 
is the short time part of the memory function, acting as a damping constant. 
$\lambda$ provides an estimate of the strength of coupling and is a 
dimensionless quantity. $\xi$ is the radius of the region of hopping, 
$\frac{(v_{0}-{\widehat {\Omega G}}({q_{m}}))}{v_{p}}$ 
gives the effect of a single hopping on the density and 
$P$ is the hopping rate per particle. Although all the parameters vary with 
temperature, the temperature dependence of 
We find that $\Omega_{0}$ and $\gamma$ have a weak temperature dependence 
which implies that the short time dynamics remain unchanged. Although,
$\lambda$ also shows a weak temperature dependence but a small variation in 
$\lambda$ leads to a substantial change in the dynamics. 
The hopping rate varies strongly with temperature.}

\begin{center}
\begin{tabular}{|c|c|c|c|c|c|c|} \hline 
&       &         &           &         &       &      \\
T (K)        &  $\Omega_{0}^{2}(ps^{-2})$ &  
$\gamma(ps^{-1})$ &$\lambda$  & $\xi/a$ & 
$\frac{(v_{0}-{\widehat {\Omega G}}({q_{m}}))}{v_{p}}$ & P ($ps^{-1}$)  \\    
	  &           &        &    &         &      &       \\ \hline
  270   & 1.353 & 1.163 & 0.94   &  1.89  & 1.79   & 3.09 $\times 10^{-4}$  \\  
  256   & 1.268 & 1.126 & 1.01   &  2.32  &3.47   & 3.2 $\times 10^{-7}$  \\  

  247   & 1.216 & 1.102 & 1.05  &  2.62  & 5.12  & 2.89 $\times 10^{-8}$  \\

&&&&&&\\\hline
\end{tabular}
\end{center} 
\end{table}

Now that we know the molecular and the thermodynamic 
parameters of the Lennard-Jones system, we can calculate  
the various parameter values in the integral equation 
that we need for a microscopic calculation,  
namely, for the MCT part, $\Omega_{0}^{2}=k_{B}Tq_{m}^{2}/m S(q_{m})$, 
$\gamma$ and $\lambda=(q_{m} A^{2}/8 \pi^{2} \rho) S(q_{m})$, 
where $A \delta(q-q_{m})=(S(q_{m})-1) $ \cite{beng}.
Instead of microscopically calculating the short time part of the memory 
kernel, $\gamma$,  we again take it to be proportional 
to $\Omega_{0}$ with the 
proportionality factor assumed to be unity (as our schematic model study). 
Gotze 
and Sjogren \cite{gotze} suggest that this proportionality factor even 
when varied between 1 and 100 does not effect the long time behavior.    
In the calculation of $\Omega_{0}$ and $\lambda$, 
$S(q_{m})$ is calculated for the above mentioned 
Lennard-Jones system.

Next for the physical quantities in the hopping part of the calculation, 
we need to calculate 
$\xi$ and $P$. These are temperature dependent quantities. For this 
calculation we use Eq.\ref{bsoft} where $T_{A}=333$ and $s_{fit}=2.65$ 
\cite{Lubwoly}. The values for all the calculated parameters are given in Table 1.

\begin{figure}
\epsfig{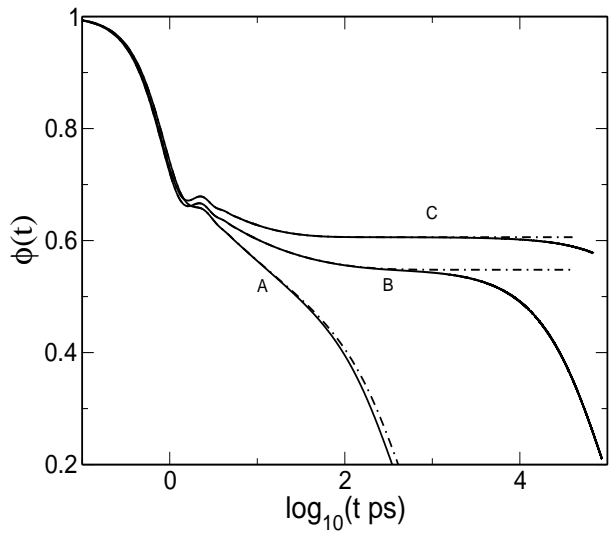}
\caption{The calculated structural relaxation for Salol 
at three state points which also includes the mode coupling transition 
temperature $T_{c}=256K$. 
Eq.\ref{schm1} is solved with the parameters given in table 1 to obtain 
$\phi(t)$. The solid lines A, B, and C correspond to $\phi(t)$ at 
temperatures 270K, 256K and 247K respectively. The dot-dashed lines are 
the plots at the same temperatures but without hopping (that is, the 
idealized MCT result).}
\end{figure}

The plot for the Salol system, obtained by using the parameters reported 
in Table 1 is given in figure 5. The plot appears to be 
quite similar to figure 1, at and below the mode coupling 
transition temperature $T_{c}=256K$. Due 
to the presence of hopping the 
structural relaxation decays from the plateau value. 
From Table 1 we find that $\Omega_{0}$ and 
$\gamma$ have a weak temperature dependence which is reflected in figure 5
in the near invariance of the short time dynamics at all 
the three temperatures. On the other hand, although
$\lambda$ shows a weak temperature dependence but a small variation in 
$\lambda$ leads to a substantial change in the dynamics. Above 
$\lambda=1$ the increase in its magnitude leads to an increase of 
the plateau value.  
The hopping rate varies strongly with temperature and has a 
stronger effect on the dynamics above $\lambda=1$ where  at lower 
temperature the plateau 
is stretched to longer times. Below $\lambda=1$ (or $T>T_{c}$), where 
$\phi(t)$ can decay completely via the diffusive channel (MCT part), 
the presence of hopping leads to a slightly 
faster decay of the structural relaxation. 

\section{Concluding Remarks}

Experiments show that the structural relaxation in a supercooled
liquid exhibits rich dynamics over many time and length scales. 
Often the physical origin of these different 
dynamics are quite disparate. To explain the dynamics of the 
liquid over the whole temperature regime under a single 
theoretical framework has been
a challenge. The random first order transition theory which 
at high temperature also contains the essential elements 
described in the perturbative mode 
coupling theory\cite{kw1987} has been successful in 
explaining the liquid dynamics 
around the laboratory glass transition temperature $T_{g}$. 
The mode coupling theory, 
on the other hand has been able to explain the high temperature dynamics extremely well. 
The theory is also known to be accurate for short times in 
the supercooled liquid below
 $T_{c}$. Below $T_{c}$ on long time scales 
the idealized MCT theory fails because of the 
exclusion of the activated hopping motion. As RFOT theory makes clear 
that hopping
motions correspond to instantons which gives rise to essential
singularities at T$_{K}$.
In this work we have proposed a unified structural description of the 
liquid, covering 
the whole temperature regime, from above $T_{c}$ to $T_{K}$.

Towards this goal we first treated the effect of hopping on the density 
fluctuation. In the theory the effect of a single particle hopping 
is connected to the neighboring density fluctuations through a propagator. 
The typical rate of a particle hopping and the spatial extent of a hopping event are
obtained from RFOT theory. After incorporating all the informations it is 
found that hopping acts as a channel for the decay of the structural 
relaxation (Eq.\ref{fhop}). Distribution of hopping times are essential 
for the non-exponentiality of the $\alpha$ relaxation \cite{lubwoly2} 
and will be incorporated in future work.
    
The total structural relaxation after incorporating the hopping motion was
calculated using two different approximate schemes.
 As in idealized MCT, the density relaxation 
is calculated self consistently with the longitudinal viscosity. 
In the first scheme (given by Eq.\ref{fqtmod}) the structural 
correlations in the time plane are taken as a simple product
of the MCT part and the hopping part.
It is found that due to the inclusion of 
the hopping motion the arrest of the 
structural relaxation at $T_{c}$, as predicted by the idealized MCT, 
disappears.
The theory with thus modified total structural relaxation predicts 
a hopping dominated decay following the MCT plateau.
In the first scheme the part of relaxation of 
$\phi(t)$ due to the diffusional motion (MCT part) 
is calculated separately (Eq.\ref{fqtmod}), but 
self-consistently with the full dynamic structure 
factor  (which now includes the hopping motion).
The theory  
predicts that hopping not only relaxes the total density correlation but 
{\it also invokes a decay of the MCT part}. This implies that hopping 
opens up continuous diffusion channels in the system. This is consistent
with the co-existence of both kinds of motions observed in computer
simulation\cite{arnab}.

It is interesting to consider the predictions of the present generalized 
theory regarding the temperature dependence of viscosity. The 
decay of the MCT part of the dynamic structure factor already predicts a 
slower rise of the viscosity with the temperature than the ideal
MCT. The interesting point to note is that the hopping rate itself
also decreases with the lowering of temperature. Thus, the dynamic 
structure factor now not only has a slow exponential decay 
in the long time,  but the plateau also gets stretched to longer times. 
Both the slow exponential 
relaxation at long times and the stretching of the plateau 
at 'intermediate times' contribute to the 
increase of the viscosity. However, the resulting temperature dependence
can be quite different from the prediction of the ideal MCT. In particular,
viscosity is dominated by the power law part at low supercooling
but at large
supercooling even the duration of the plateau and the power law
decay are determined mostly by the hopping rate.

{\bf ACKNOWLEDGEMENT}~
 This work was supported in parts from NSF (USA) and DST (India).

\end{document}